\documentclass[twocolumn, longbib]{aastex701}
\shorttitle{Raman Spectroscopy of Enceladus Salt Deposits}
\shortauthors{Takeshita et al.}

\begin{document}

\title{Raman Spectroscopy of Salt Deposits from the Simulated Subsurface Ocean of Enceladus}

\author{Jun Takeshita}
\affiliation{The University of Tokyo, Department of Earth and Planetary Science, 7-3-1 Hongo, Bunkyo, Tokyo 113-0033, Japan}
\email{takeshita-jun878@g.ecc.u-tokyo.ac.jp}

\author[orcid=0000-0003-2749-2204]{Yuichiro Cho}
\affiliation{The University of Tokyo, Department of Earth and Planetary Science, 7-3-1 Hongo, Bunkyo, Tokyo 113-0033, Japan}
\email{cho@eps.s.u-tokyo.ac.jp}

\author[orcid=0000-0002-5653-5982]{Haruhisa Tabata}
\affiliation{Institute of Space and Astronautical Science (ISAS), Japan Aerospace Exploration Agency (JAXA), 3-1-1 Yoshinodai, Sagamihara, Kanagawa 252-5210, Japan}
\email{tabata.haruhisa@jaxa.jp}

\author[orcid=0000-0001-7860-2341]{Yoshio Takahashi}
\affiliation{The University of Tokyo, Department of Earth and Planetary Science, 7-3-1 Hongo, Bunkyo, Tokyo 113-0033, Japan}
\email{ytakaha@eps.s.u-tokyo.ac.jp}

\author[orcid=0000-0001-6423-0698]{Daigo Shoji}
\affiliation{The University of Tokyo, Department of Earth and Planetary Science, 7-3-1 Hongo, Bunkyo, Tokyo 113-0033, Japan}
\email{shoji.daigo@eps.s.u-tokyo.ac.jp}

\author[orcid=0000-0001-6076-3614]{Seiji Sugita}
\affiliation{The University of Tokyo, Department of Earth and Planetary Science, 7-3-1 Hongo, Bunkyo, Tokyo 113-0033, Japan}
\email{sugita@eps.s.u-tokyo.ac.jp}


\begin{abstract}

Saturn’s ice-covered moon Enceladus may host a subsurface ocean with biologically relevant chemistry. Plumes released from this ocean preserve information on its chemical state, and previous analyses suggest weakly to strongly alkaline pH (approximately pH 8-12). Constraining the pH requires identification of pH-sensitive minerals in plume deposits. Several analytical techniques could provide such mineralogical information, but few are practical for deployment on planetary missions. Raman spectrometers, which have recently advanced for in situ exploration and have been incorporated into flight instruments, offer a feasible approach for mineral identification on icy moons. However, their applicability to pH estimation from plume-derived minerals has not been investigated. In this study, we evaluate whether Raman measurements of plume particles deposited on the surface of Enceladus can be used to distinguish between weakly and strongly alkaline subsurface ocean models. Fluids with pH values of 9 and 11 were frozen under vacuum conditions analogous to those on Enceladus. The resulting salt deposits were then analyzed using a flight-like Raman spectrometer. The Raman spectra show pH-dependent carbonate precipitation: NaHCO$_3$ and Na$_2$CO$_3$ peaks were detected at pH 9, whereas only Na$_2$CO$_3$ peaks were detected at pH 11. These findings demonstrate that Raman spectroscopy can distinguish pH-dependent carbonate phases. This capability allows us to constrain whether the pH of the subsurface ocean is weakly alkaline or strongly alkaline, which is a key parameter for assessing its chemical evolution and potential habitability.

\end{abstract}



\section{Introduction} \label{sec:intro}
Enceladus is one of Saturn's satellites and is covered by ice. It erupts large-scale cryovolcanic plumes from aligned vent structures, termed ``Tiger Stripes'', in its southern polar region \citep{2006Sci...311.1401S,2006Sci...311.1393P}. These plumes are considered to originate from the global subsurface ocean \citep{2011Natur.474..620P,2016Icar..264...37T,2017NatAs...1..841C} and are the main source of Saturn's E-ring \citep{2008Natur.451..685S,2008Icar..193..420K}. The Cosmic Dust Analyzer (CDA) onboard Cassini analyzed the composition of E-ring particles. \citet{2009Natur.459.1098P} classified these particles into three types based on the mass spectra: TI (pure water ice), TII (organic- or silicate-rich ice), and TIII (salt-rich ice). They revealed that the ``TIII'' particles in the E-ring were made by the rapid freezing of liquid spray \citep{2009Natur.459.1098P} and best represent the composition of the subsurface ocean \citep{2011Natur.474..620P,2018eims.book..129P}. TIII particles make up $\sim$6\% of the E-ring ice particles. These particles have high concentrations of Na salts, which are interpreted to be frozen submicron-size droplets that were erupted directly from a subsurface salt-rich ocean. Their presence provides strong evidence that liquid water currently exists beneath the icy shell of Enceladus. The CDA data indicate that the subsurface ocean hosts an alkaline fluid consisting mainly of Na, chlorides, and carbonates \citep{2015GeCoA.162..202G,2017Sci...356..155W,2020GeRL...4786484G}.

The ocean pH can affect the solubility of ions and even its habitability, and is therefore important for understanding the subsurface ocean and might provide insights into the interior of Enceladus \citep{2021PSJ.....2..132C}. The pH of the subsurface ocean on Enceladus is constrained to be alkaline (pH $>$ 7), but different studies have proposed pH values ranging from weakly alkaline (pH = 8--9) to strongly alkaline (pH = 10--12). For example, \citet{2022PSJ.....3..191F} estimated a pH of 7.95--9.05 based on numerical modeling of plume fractionation processes, while \citet{2025Icar..42616717G} estimated a pH of 10.1--11.6 based on geochemical modeling of phosphate speciation observed in plume ice grains. The dissolved species of carbonate and phosphate ions change significantly between pH = 9 and pH = 11. At pH = 9, HCO$_3^-$ is the predominant species, while almost all carbonate is CO$_3^{2-}$ at pH = 11. Similarly, HPO$_4^{2-}$ and H$_2$PO$_4^-$ are dominant at pH = 9, while PO$_4^{3-}$ appears at pH = 11 \citep{2021JGRE..12660516F,2025Icar..42616717G}. Furthermore, whether the pH is weakly or strongly alkaline changes the amount of chemical energy available for life in the subsurface ocean. At pH = 11, strong serpentinization reactions occur continuously, producing large amounts of H$_2$. The reaction between H$_2$ and CO$_2$ produces methane, which yields the energy used by life \citep{2023FrMic..1457597S}. At pH = 9, limited serpentinization leads to little H$_2$. The conditions in the subsurface ocean have a strong effect on the sustainability of life \citep{2023FrMic..1457597S}.

Salt deposits on the surface of Enceladus are formed near the Tiger Stripes region as water sublimates from the plume-derived ice, and should reflect the subsurface ocean pH \citep{2009Natur.459.1098P,2015GeCoA.162..202G}. To investigate the pH-dependence of relevant materials on Enceladus, \citet{2021JGRE..12660516F} prepared simulated Enceladus subsurface ocean solutions at pH = 9 and 11, froze them, sublimated the ice, and identified the salt species and texture by cryo-scanning electron microscopy (SEM), cooling-stage optical microscopy, and powder X-ray diffraction (XRD) analysis. They found that thermonatrite (Na$_2$CO$_3 \cdot$H$_2$O) was absent at pH = 9, but present at pH = 11; trona (Na$_2$CO$_3 \cdot$NaHCO$_3 \cdot$2H$_2$O) was detected at pH = 9, but only weakly at pH = 11; and nahcolite (NaHCO$_3$) was detected at pH = 9 and not at pH = 11. These results highlight the influence of pH on the mineralogy.

Proposed lander missions to Enceladus \citep[e.g.,][]{2020FrASS...7...26N,2021PSJ.....2..100C,2021PSJ.....2...77M} are prime opportunities for the on-site analysis of such salt deposits. For example, the Enceladus \textit{Orbilander} concept plans to carry several instruments, including a mass spectrometer, electrochemical sensors, and context imager for in situ analyses of surface materials of Enceladus \citep{2021PSJ.....2...77M}. Similarly, the \textit{SILENUS} concept involves the delivery of penetrators equipped with a seismometer and context imager to observe the surface morphology at multiple south polar sites \citep{2022FrASS...995941N}. Other lander mission concepts would carry the Ion Microscope and Mass Spectrometer Suite \citep[\textit{ETNA};][]{2022FrASS...928357D}, a nanopore sensor, a multiplex antibody microarray, a microscope system with flow cytometry, and a gas chromatography mass spectrometer for lipid biomarker analysis \citep[\textit{EnEx};][]{2015AcAau.106...63K}. For the Enceladus lander missions, determining the mineralogical composition of the material that originates from its interior is critical. Raman spectroscopy can provide important mineral compositional information, as successfully demonstrated in Mars missions \citep{2021SSRv..217....4W,2023Natur.619..724S}. Unlike mass spectrometry, which primarily determines elemental and isotopic composition, Raman spectroscopy can directly identify specific mineral phases (e.g., Na$_2$CO$_3$ vs. NaHCO$_3$) without the need for sample processing. This capability is useful for constraining geochemical parameters such as pH, which controls the specific salt phases precipitated from the subsurface ocean. Although in situ exploration of Europa with Raman spectroscopy has also been proposed \citep{2019LPI....50.2992P,2020JRS....51.1782S}, this technique has not been evaluated specifically as a tool to constrain the pH of the subsurface ocean of Enceladus by mineralogical analysis.

In this study, we investigated whether Raman spectroscopy could be used to identify minerals specific to the possible pH conditions of the subsurface ocean of Enceladus. We used flight-like instruments to demonstrate the feasibility of Raman spectroscopy for future Enceladus missions.


\section{Methods} \label{sec:methods}
\subsection{Preparation of salt samples} \label{subsec:preparation}

\begin{deluxetable*}{lcccc}[ht!]
\tablecaption{Compositions of the brines used to simulate the subsurface ocean of Enceladus \citep[based on][]{2021JGRE..12660516F} \label{tab:brine_composition}}
\tablehead{
    \colhead{} & \multicolumn{2}{c}{pH = 9} & \multicolumn{2}{c}{pH = 11} \\
    \cline{2-3} \cline{4-5}
    \colhead{} & \colhead{Weight of reagent} & \colhead{Concentration} & \colhead{Weight of reagent} & \colhead{Concentration} \\
    \colhead{Species} & \colhead{[g/100 g H$_2$O]} & \colhead{[mol/L]} & \colhead{[g/100 g H$_2$O]} & \colhead{[mol/L]}
}
\startdata
NaCl & $1.3 \times 10^{1}$ & $2.2 \times 10^{0}$ & $1.2 \times 10^{1}$ & $2.0 \times 10^{0}$ \\
Na$_2$CO$_3$ & $8.2 \times 10^{-1}$ & $7.8 \times 10^{-2}$ & $7.4 \times 10^{0}$ & $7.0 \times 10^{-1}$ \\
NaHCO$_3$ & $5.2 \times 10^{0}$ & $6.2 \times 10^{-1}$ & --- & --- \\
NH$_4$Cl & $5.3 \times 10^{-2}$ & $9.9 \times 10^{-3}$ & $5.3 \times 10^{-2}$ & $9.9 \times 10^{-3}$ \\
KCl & $2.5 \times 10^{-1}$ & $3.3 \times 10^{-2}$ & $2.5 \times 10^{-1}$ & $3.3 \times 10^{-2}$ \\
Na$_2$SiO$_3$ & $4.1 \times 10^{-1}$ & $3.8 \times 10^{-2}$ & $4.1 \times 10^{-1}$ & $3.8 \times 10^{-2}$ \\
\enddata
\end{deluxetable*}

We prepared two types of brine samples with pH = 9 and 11. These pH values are close to the upper and lower limits of the predicted pH of the subsurface ocean of Enceladus \citep{2021JGRE..12660516F}, where organic matter is expected to be stably synthesized \citep{2024ApJ...971...51L}. The solutes of these fluids were set following the results of \citet{2021JGRE..12660516F} (Table 1). Solutions at pH = 9 and 11 were prepared by mixing reagents according to the weights in Table 1. For the pH = 9 solution, NaCl, Na$_2$CO$_3$, NaHCO$_3$, NH$_4$Cl, KCl, and Na$_2$SiO$_3$ were combined and dissolved in 100~g of distilled water. For the pH = 11 solution, NaCl, Na$_2$CO$_3$, NH$_4$Cl, KCl, and Na$_2$SiO$_3$ were mixed and dissolved in 100~g of distilled water.

\begin{figure*}[ht!]
\plotone{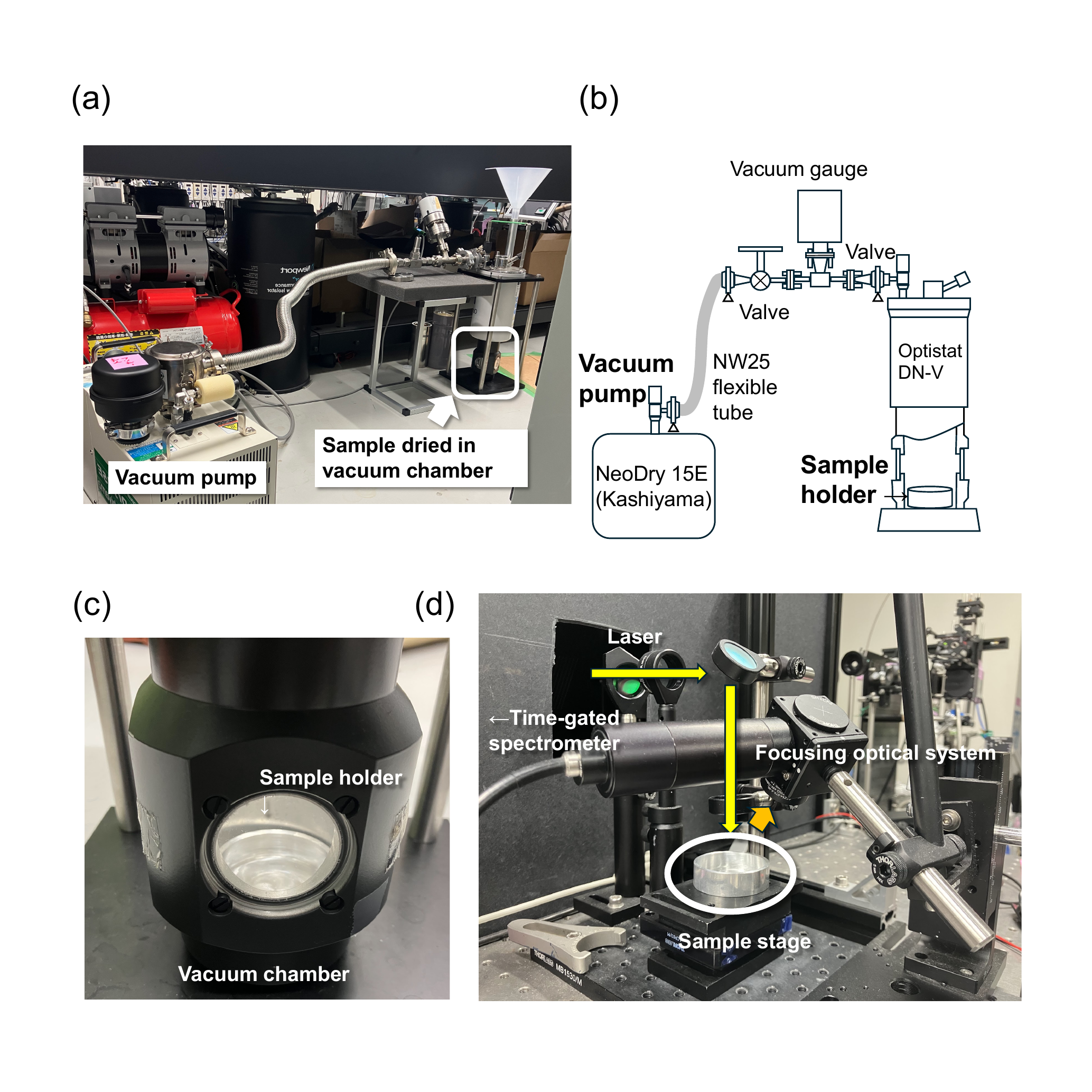}
\caption{Cooling experimental apparatus. (a) Photograph and (b) schematic diagram of the dry freezing system. (c) The sample chamber with a sample (i.e., a frozen solution) in the chamber. Above this is a cooler filled with liquid nitrogen. (d) Photograph of the Raman spectrometer. \label{fig:apparatus}}
\end{figure*}

\begin{figure*}[ht!]
\plotone{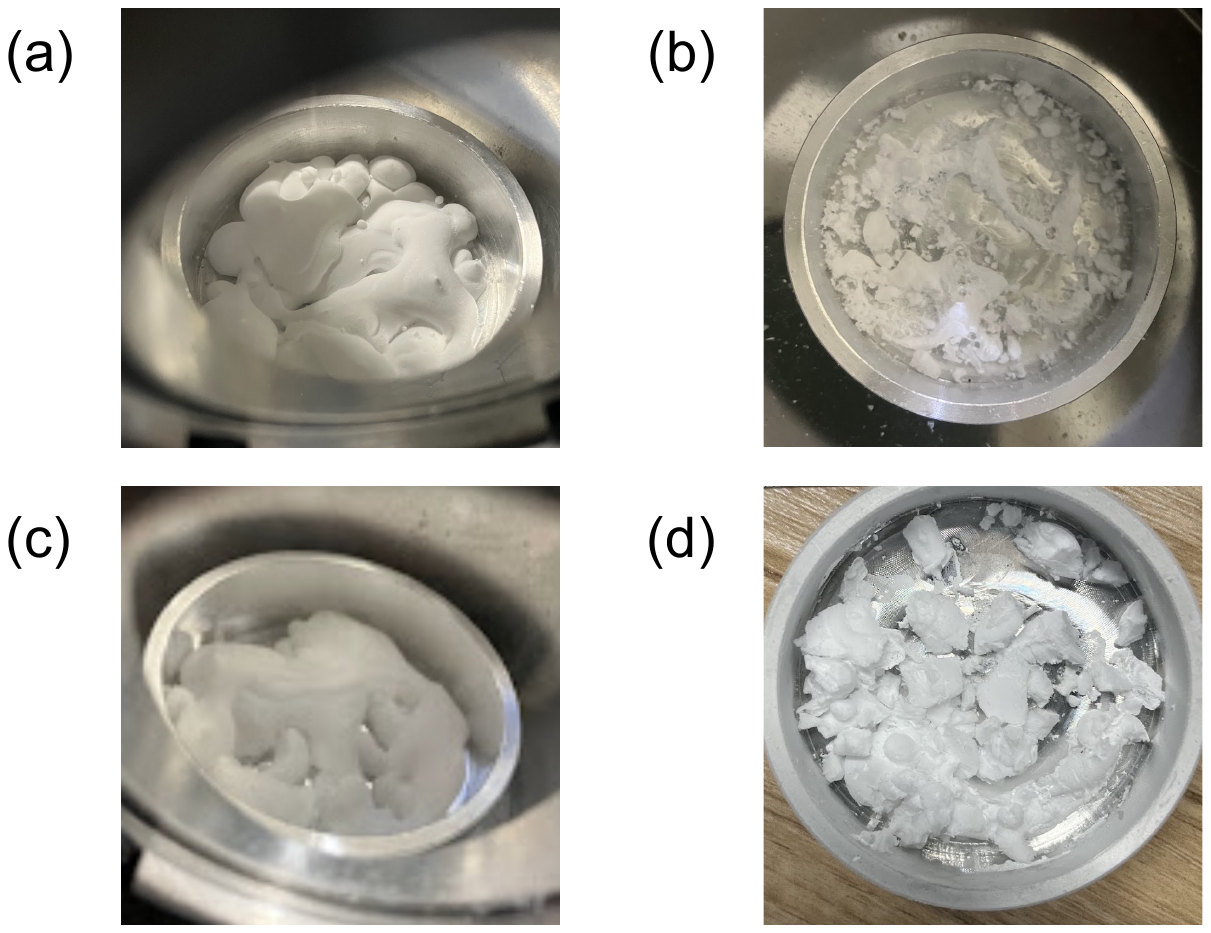}
\caption{Photographs of the samples at (a) pH = 9 immediately after starting vacuum pumping and (b) pH = 9 after 18~h of vacuum pumping, (c) pH = 11 immediately after starting vacuum pumping, and (d) pH = 11 after 18~h of vacuum pumping. The sample holder is 5~cm in diameter. \label{fig:samples}}
\end{figure*}

A volume of 5.0~mL of the prepared pH = 9 or 11 solutions was pipetted onto a metallic holder. The brine solutions were then frozen by placing the metallic holder in a chamber pre-cooled to $\sim$100~K by liquid nitrogen. The chamber was then evacuated with a dry pump to evaporate the water in the brine solutions, producing dried salt samples. We also prepared powder samples without dry freezing to evaluate the differences between the salts that were experimentally precipitated from the liquid and the standard powdered samples, such as the effects of uneven mixing and crystallization. The powder compositions (Table 1) were NaCl, Na$_2$CO$_3$, NaHCO$_3$, NH$_4$Cl, KCl, and Na$_2$SiO$_3 \cdot$9H$_2$O. The NaCl, Na$_2$CO$_3$, NaHCO$_3$, and KCl used in this study were obtained from Wako Pure Chemical Industries Limited. First-grade quality reagents, with a purity of $>$99.5\%, were used. However, for the Na$_2$SiO$_3 \cdot$9H$_2$O, the purity was $>$98.0\%. Although the plume contains a wide variety of C compounds \citep{2024NatAs...8..164P}, no organic substances were added in this experiment.

\subsection{Dry freezing of sample solutions} \label{subsec:freezing}

In each cooling experiment, one metallic sample holder was placed in the vacuum chamber (Fig. \ref{fig:apparatus}a--b; Oxford Instruments Optistat DN-V) to precipitate salts from the frozen brine samples. The pressure in the vacuum chamber was held at 1--10~Pa, which was comparable to the environment around Tiger Stripes. Each sample was left in the chamber for 18~h to produce precipitated salts after water sublimation. The sample was gradually annealed to room temperature by heat conduction from the bottom of the sample holder. After 18~h of dry cooling, the samples were removed from the vacuum chamber and immediately analyzed by a Raman spectrometer at room temperature and pressure. These experimental conditions were designed to reproduce the mineralogical state of plume-derived material deposited in the warm regions near Tiger Stripes, after the plumes are erupted into space and gently annealed once settled on the surface.

Figure \ref{fig:samples}a and c shows images of the samples produced by the experiments. The droplets were placed by pipette into the frozen sample holder and became ice grains containing salt with particle sizes of several millimeters to several centimeters. The sample images after 18~h under low-temperature vacuum conditions are shown in Fig. \ref{fig:samples}b and d. The samples are dry solids with a particle size of several millimeters. The samples were analyzed by Raman spectroscopy in their intact form, with no crushing or additional treatment (see section \ref{subsec:raman}).

\begin{deluxetable*}{lll}[ht!]
\tablecaption{Comparison of the key specifications of the Raman spectroscopic systems used in this study and the SuperCam \label{tab:raman_spec}}
\tablehead{
    \colhead{Parameter} & \colhead{This Study} & \colhead{SuperCam \citep{2021SSRv..217....4W}}
}
\startdata
Operational Concept & Standoff (Short-range) & Remote Sensing (Long-range) \\
Standoff Distance & 40~mm & Up to $\sim$7~m \\
Spectral Resolution (FWHM) & 15--20~cm$^{-1}$ & $\le$12~cm$^{-1}$ \\
Spectrometer & SpectraPro 300i & High-efficiency transmission spectrometer \\
Excitation Laser & 532~nm Nd:YAG & 532~nm Nd:YAG (frequency-doubled) \\
Laser Energy (532~nm) & $\sim$8~mJ/pulse & $\sim$8~mJ/pulse \\
Laser spot diameter & $\sim$3~mm & $\sim$5~mm (when standoff distance is 7~m) \\
ICCD Gate Width & 10~ns & 100~ns \\
ICCD Gate Delay & 8.361~$\mu$s & $\sim$650~ns \\
Pulse Accumulation & 1000~pulses & 100--200~pulses \\
\enddata
\end{deluxetable*}

\subsection{Raman spectroscopy} \label{subsec:raman}

We constructed a Raman spectroscopic system to simulate the specifications of equipment required for future exploration of icy satellites. The specifications of this system (Table \ref{tab:raman_spec}) are designed to be similar to those of the SuperCam instrument on the NASA Perseverance Rover \citep{2021SSRv..217....4W,2021SSRv..217...47M}. Raman spectra were acquired with a SpectraPro 300i spectrometer equipped with a PI-MAX 1K RB-133 ICCD camera and a Nd:YAG laser ($\lambda = 532$~nm) for excitation. The spectral window was 539.2--820.1~nm (corresponding to Raman shifts of 250--6600~cm$^{-1}$) with a 150~lpm grating centered at 680~nm; and 541.9--571.9~nm (343--1312~cm$^{-1}$), with a 1200~lpm grating centered at 557~nm. The wavelength resolution was 0.28~nm/pixel ($\approx$4~cm$^{-1}$). To ensure the exposure occurred instantly after a laser shot, the ICCD exposure time and gate delay were set to 10~ns and 8.361~$\mu$s, respectively. The laser was operated at 10~Hz, with an 8~mJ pulse energy and 3~mm beam diameter, and each spectrum was accumulated over 1000 laser pulses.

Although based on the SuperCam instrument, our system has distinct operational parameters. Firstly, all spectra were acquired at a fixed standoff distance of 40~mm from the sample. This near-contact configuration differs significantly from the remote analysis design of SuperCam, which can acquire data from targets at distances of up to $\sim$7~m. Secondly, the effective instrumental spectral resolution, defined by the full-width-at-half-maximum (FWHM), is 15--20~cm$^{-1}$. This is wider than SuperCam in high-resolution mode, which has a FWHM of $\le$12~cm$^{-1}$ in the fingerprint region. For the analysis, the laser was applied directly to the sample in a sample holder after it was removed from the chamber, and the measurements were undertaken at one point on each sample (Fig. \ref{fig:samples}b and d).

\section{Results} \label{sec:results}
\begin{figure*}[ht!]
\plotone{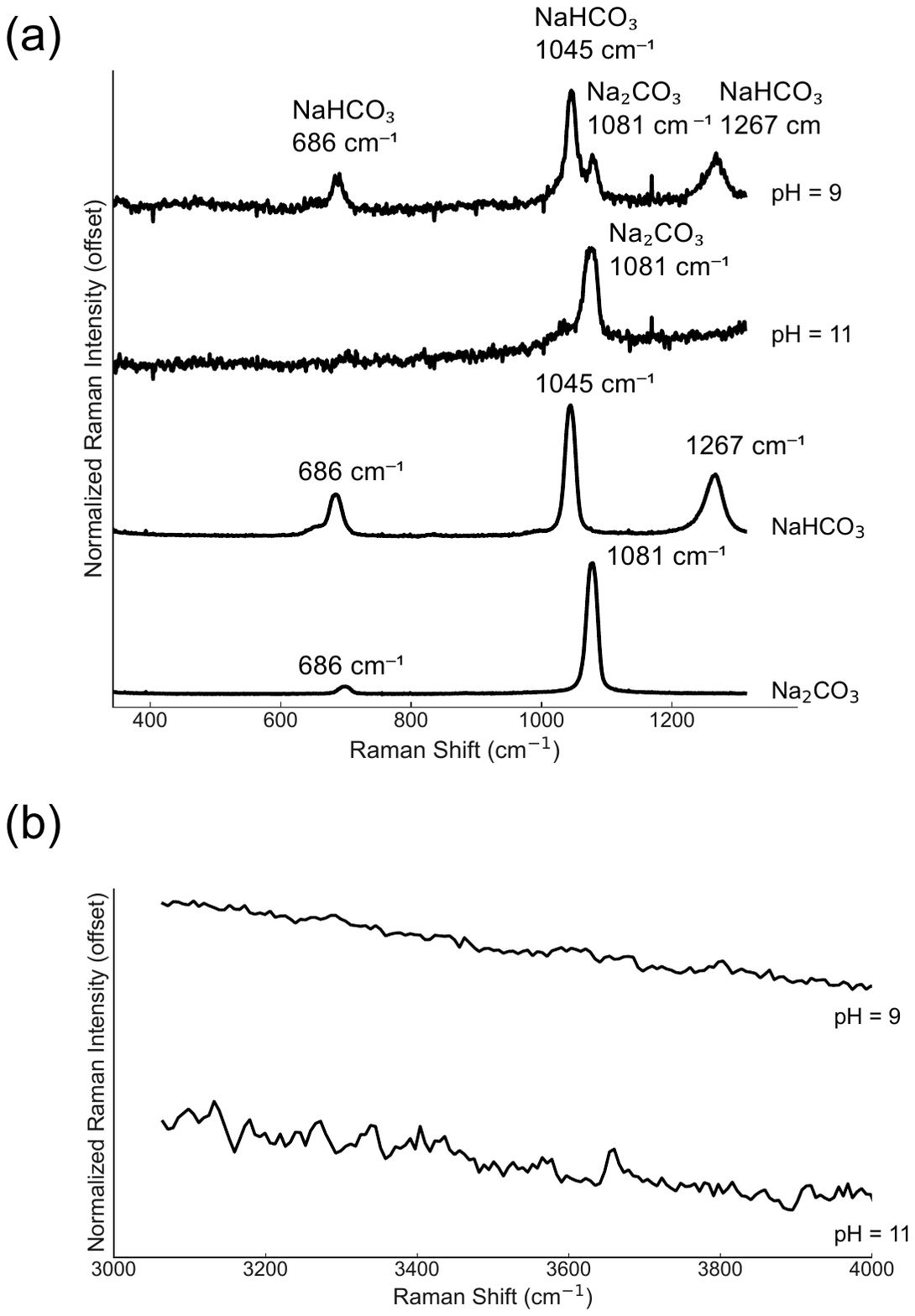}
\caption{Raman spectra of samples under different pH conditions. (a) Spectra in the 400–1300~cm$^{-1}$ range; measured spectra of standard $\mathrm{NaHCO_{3}}$ and $\mathrm{Na_{2}CO_{3}}$ powders are shown at the bottom for reference. Raman peaks observed at approximately 686, 1045, and 1267 cm$^{-1}$ are assigned to vibrational modes of the bicarbonate ion (HCO$_3^-$).Raman peaks observed at approximately 686 and 1081 cm$^{-1}$ is assigned to vibrational modes of the carbonate ion (CO$_3^{2-}$).
(b) Spectra in the 3200-4000~cm$^{-1}$ range.
 \label{fig:spectra}}
\end{figure*}

Raman spectra of the samples at the original concentration for each pH are shown in Fig.~\ref{fig:spectra}a. 
Peaks at 686, 1045, and 1267~cm$^{-1}$ were detected at pH = 9. 
When compared with the Raman spectra of the powdered reagents shown in Fig.~\ref{fig:spectra}a, these peaks appear to be attributable to NaHCO$_3$.
These Raman peaks at approximately 686, 1045, and 1267~cm$^{-1}$ are assigned to vibrational modes of the bicarbonate ion (HCO$_3^-$).
Notably, these peaks were not observed at pH = 11. 
At pH = 11, a peak at 1081~cm$^{-1}$ was detected, which is attributable to Na$_2$CO$_3$. 
This Raman band at approximately 1081~cm$^{-1}$ is assigned to vibrational modes of the carbonate ion (CO$_3^{2-}$).
This 1081~cm$^{-1}$ peak was also observed at pH = 9.

The spectra in Fig. \ref{fig:spectra}a display a Raman shift from 400 to 1300~cm$^{-1}$, whereas Fig. \ref{fig:spectra}b covers a range from 3200 to 4000~cm$^{-1}$. In the spectra in Fig. \ref{fig:spectra}b, no peaks corresponding to water were detected around the 3400~cm$^{-1}$ range, indicating that the water in the sample sublimated during the freeze drying (Fig. \ref{fig:apparatus}b) and no deliquescence occurred during the measurements.

The detection of distinct carbonate phases (NaHCO$_3$ vs Na$_2$CO$_3$) from the frozen-dried samples is significant because it suggests that the chemical signature of the original ocean pH is preserved in the mineralogy of the salt deposits. Furthermore, clear Raman peaks were obtained even from samples with realistic salt concentrations comparable to the subsurface ocean and without any concentration or purification processes. This suggests that a flight-like Raman spectrometer can directly distinguish the pH environment of the subsurface ocean solely by analyzing surface salt deposits.

The results of our experiments are consistent with the findings of \citet{2021JGRE..12660516F}, who analyzed their samples by XRD, SEM, and energy dispersive spectrometry. In their study, Na$_2$CO$_3 \cdot$NaHCO$_3 \cdot$2H$_2$O (trona) and NaHCO$_3$ (nahcolite) were identified as the dominant phases at pH = 9, while Na$_2$CO$_3 \cdot$H$_2$O (thermonatrite) was mainly detected at pH = 11, along with some NaHCO$_3$. Our experiments demonstrate that Raman spectroscopy yields result consistent with this previous study, confirming its effectiveness in detecting these mineral phases. In the next section, we discuss the limitations of our experiments, caveats regarding the actual measurements, and the implications of our results.

\section{Discussion} \label{sec:discussion}

\subsection{Relevance of Experimental Conditions to the Enceladus Surface Environment} \label{subsec:limitations}

The results presented in this study are based on a fixed freezing rate. Although the freezing velocity of a plume depends on eruptive processes that are poorly understood \citep{2024JGRE..12908070M}, variations in the freezing rate could affect the spectra, such as by amorphization. In fact, \citet{2023PSJ.....4..156V} measured a sample of rapidly cooled brine containing salt and organic matter using Raman spectroscopy. Only peaks from water were observed and no peaks from salt were detected. However, when the temperature of the sample was increased from 110 to 220~K, peaks attributable to NaCl$\cdot$2H$_2$O (hydrohalite) and amino acids (glycine) appeared. This was because the peaks were hidden by broad peaks from amorphous components. In our experiments, the sample was frozen and dried, and then measured at room temperature, which would have a similar effect as annealing, allowing for the detection of Raman peaks. On Enceladus, crystalline ice exists near the relatively hot Tiger Stripes region, while amorphous ice exists in the cooler surrounding areas \citep{2006Sci...311.1425B}. In the present study, the measured samples correspond to particles that were flash-frozen during plume ejection and subsequently deposited and annealed near the Tiger Stripes region.

The Raman measurements in this study were conducted at an ambient temperature and pressure. To fully replicate the environmental conditions on Enceladus, future experiments should include Raman measurements at low temperatures and pressures. However, the distinction between the NaHCO$_3$ and Na$_2$CO$_3$ identified in this study is based mainly the pH of the parent fluid, not the measurement temperature. In fact, measurements of single trona crystals at 100--340~K showed no systematic change in the Raman shift \citep{2014AmMin..99.1973O}. When the phase is the same, the Raman shift due to the temperature difference ($\sim$180 and 300~K) would be limited to a small shift and line width change of a few cm$^{-1}$. Based on these findings, the results of this study adequately represent those that would be obtained when such measurements would actually be conducted near the south pole of Enceladus.

\begin{figure*}[ht!]
\plotone{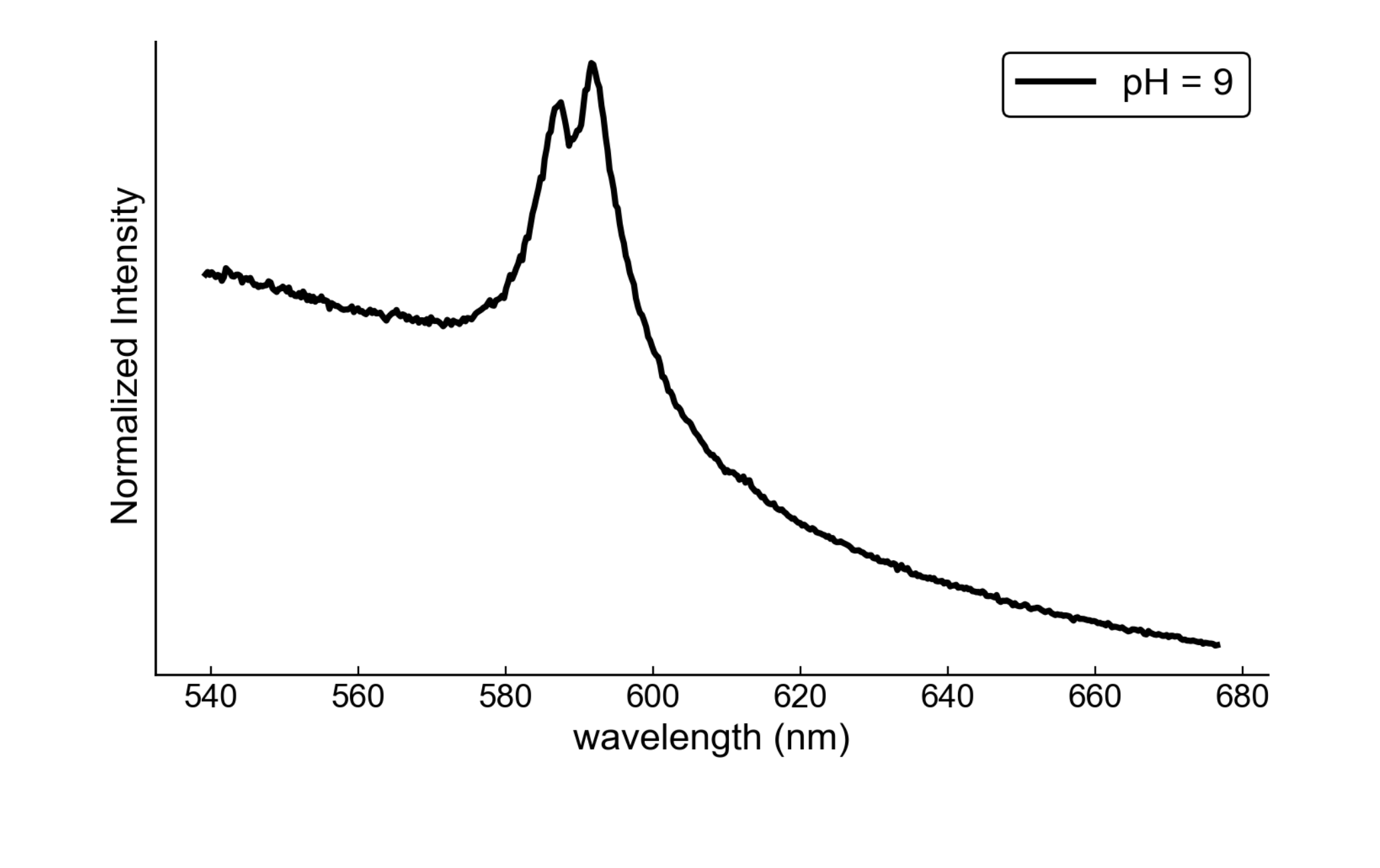}
\caption{Example of a spectrum with the emission lines of Na atoms. Accumulation count = 100 times, grating = 150~lpm, and exposure time = 10~ns. \label{fig:libs}}
\end{figure*}

We also note that the high-intensity pulsed laser could cause laser-induced breakdown spectroscopy, when the sample distance is too close to the laser focus length (Fig. \ref{fig:libs}). The wavelengths of the two peaks are 587.5 and 591.6~nm, respectively. Given that the separation of the two peaks is wider than the DI (589.592~nm) and DII (588.995~nm) lines of the Na atomic emission \citep{2013JAAS...28...92K}, this splitting would result from strong Na self-absorption in the laser-induced vapor. A continuum emission beneath the strong Na emission lines is also evident. Given that the intensities of such laser-induced breakdown spectroscopic signals are much greater than those of Raman scattering, it is necessary to adjust the focus position during the actual observations to avoid laser-induced breakdown of the samples.

\subsection{Formation of Surface Salt Deposits and Implications for the Subsurface Ocean} \label{subsec:timescale}

\begin{deluxetable*}{lll}[ht!]
\tablecaption{Parameters used in our calculations \label{tab:sublimation_params}}
\tablehead{
    \colhead{Parameter} & \colhead{Value} & \colhead{Note}
}
\startdata
Temperature $T$ [K] & 140--200 & Based on \citet{2009sfch.book..683S}; \citet{2018eims.book..129P} \\
Radius of grains $r$ [$\mu$m] & 50 & Based on \citet{2018eims.book..129P} \\
Sublimation coefficient $\alpha$ & 1 & Assuming ideal sublimation \\
Molar mass of water $M$ [kg/mol] & 0.01801528 & \\
Gas constant $R$ [J/mol/K] & 8.314 & \\
Density of ice $\rho_{ice}$ [kg/m$^3$] & 917 & \\
\enddata
\end{deluxetable*}

In this study, we analyzed salt deposits derived from the frozen droplets of sample solutions (Table \ref{tab:brine_composition}). To evaluate whether salt-containing water ice would actually sublimate into dry salt when a plume is erupted and deposited on the surface of Enceladus, we calculated how long it would take for the ice in the plume to sublimate in the environment around Tiger Stripes, assuming that it was deposited on the ground after the plume erupted. The parameters used in the following discussion are listed in Table \ref{tab:sublimation_params}. We used the Hertz--Knudsen (HK) equation to estimate the sublimation rates. The HK equation is used in cases of plume evaporation on icy moons \citep[e.g.,][]{2020FrASS...7...14P}. Following this method, the mass flux $J$ of sublimating ice is given by

\begin{equation}
J = \alpha \cdot \frac{P_{sat}M}{\sqrt{2\pi RT}} \label{eq:hk},
\end{equation}

where $\alpha$ is the sublimation coefficient (assumed to be 1), $P_{sat}$ is the saturation vapor pressure of ice at temperature $T$, $M$ is the molar mass of water, and $R$ is the gas constant. To estimate the duration of sublimation, we considered a spherical ice particle of radius $r$ that was deposited on the surface and exposed only in its upper hemisphere. This geometry assumes that the lower hemisphere is in direct contact with the ground and thus protected from sublimation into a vacuum, which is a plausible condition for particles settling onto the surface. Given that sublimation occurs only over the upper hemispherical surface, the effective sublimation area is half of the total surface area. The sublimation time $t$ is defined as the time taken for the particle to completely sublimate under these conditions, which is given by

\begin{equation}
t = \frac{m}{\frac{dm}{dt}} = \frac{m}{J \cdot A} = \frac{2\rho_{ice} \cdot r}{3J} \label{eq:time},
\end{equation}

where $\rho_{ice}$ is the bulk density of water ice. This equation accounts for the reduced sublimation surface due to ground contact and provides a more realistic estimate of the sublimation time of ice particles on Enceladus.

As sublimation proceeds, salts such as NaCl and Na$_2$CO$_3$ precipitate at the surface and form a coating that inhibits water vapor dissipation. This effect can be modeled by introducing a suppression factor $f$ ($0 < f \le 1$) as follows:

\begin{equation}
t_{\mathrm{coated}} = \frac{t_{\mathrm{bare}}}{f}, \label{eq:coated}
\end{equation}

where $t_{\mathrm{coated}}$ is the time required for complete mass loss in the exposed hemisphere and $t_{\mathrm{bare}}$ is the time assuming no surface coating and unrestricted sublimation. For example, under typical Enceladus conditions ($T = 180$~K, $r = 50$~$\mu$m, and assuming $f = 0.5$) \citep{2009sfch.book..683S,2018eims.book..129P}, we estimate sublimation times of ca. 17~h for salt-coated ice. Note that the saturation vapor pressure was calculated using eq. 7 of \citet{2005QJRMS.131.1539M}, as follows:

\begin{equation}
\ln P_{sat} = 9.550426 - \frac{5723.265}{T} + 3.53068 \ln T - 0.00728332 T \label{eq:psat}.
\end{equation}

\begin{figure*}[ht!]
\plotone{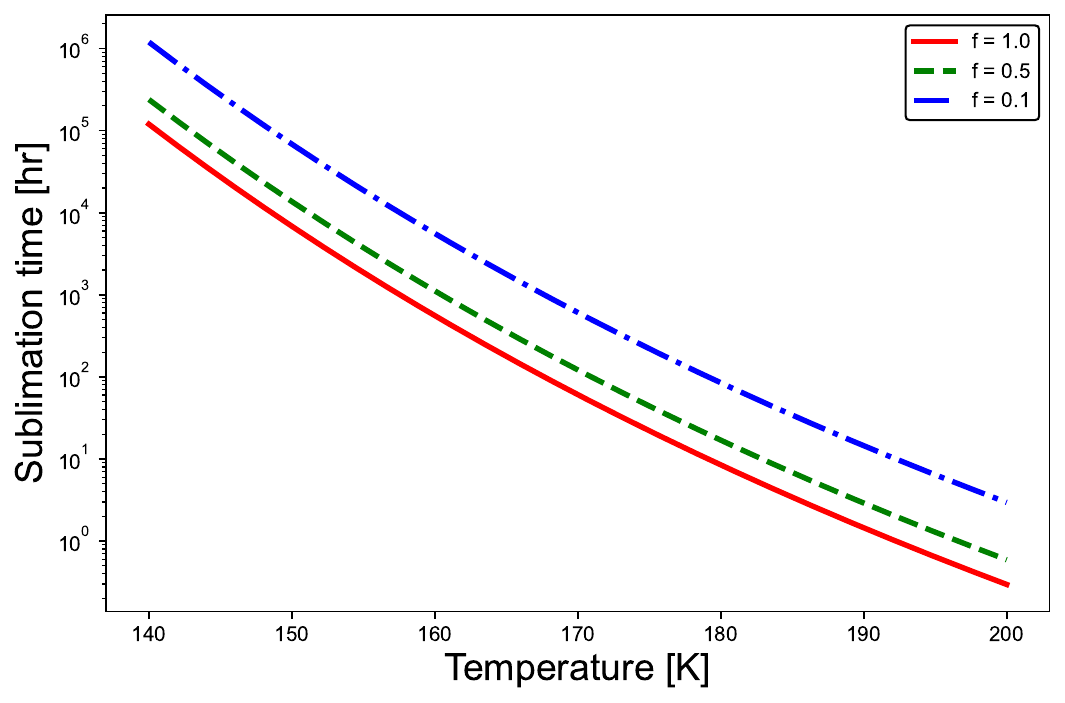}
\caption{Sublimation time as a function of temperature for different coating suppression factors ($f = 1.0$, $0.5$, and $0.1$). Salt coatings substantially increase the survival time of icy grains under Enceladus-like conditions. \label{fig:sublimation}}
\end{figure*}

We further calculated the sublimation times as a function of temperature from 140 to 190~K, assuming three different coating efficiencies ($f = 1.0$, $0.5$, and $0.1$). The sublimation time decreases with increasing temperature due to the exponential increase in $P_{sat}$ (Fig. \ref{fig:sublimation}). As the value of $f$ approaches 1 (i.e., the effect of the salt coating on the surface in suppressing water evaporation increases), the time required for sublimation increases. Ice can persist for years at 140~K, but disappears within hours at 180~K. This supports the hypothesis that the plume contains salt and that ice sublimes and turns into dry salt after deposition. In the Tiger Stripes region, where temperatures reach $\sim$180~K, evaporation of water could easily occur on a timescale of a few hours.

It is also important to consider the pH consistency, as the pH of the subsurface ocean does not necessarily match the pH of the surface material ejected through various processes. It is necessary to track the pH changes during eruption processes. To relate the pH recorded by plume-derived salts to the subsurface ocean pH, \citet{2025Icar..42616717G} accounted for CO$_2$ degassing during eruption. Because CO$_2$ behaves as a weak acid in water, its removal drives changes in dissolved ions that increase the pH. By constraining this process with a degassing model, \citet{2025Icar..42616717G} found that in a realistic scenario the total pH increase during outgassing is minimal ($<0.2$), implying that the pH measured in plume salt solutions should be representative of the subsurface ocean. Therefore, we can use the pH of the plume-derived salts to infer the subsurface ocean pH.

\section{Conclusions} \label{sec:conclusions}

We conducted laboratory experiments that simulated the subsurface ocean of Enceladus by preparing two fluid compositions with pH = 9 and 11. Using a freeze-drying apparatus, we froze the simulated ocean samples and sublimated the water ice, leaving only the salt deposits for Raman spectroscopic measurements. As a result, peaks corresponding to different mineral precipitates were detected depending on the pH. At pH = 9, peaks at 686, 1045, and 1267~cm$^{-1}$ were detected, which are consistent with NaHCO$_3$. At both pH = 9 and 11, a clear peak appeared at 1081~cm$^{-1}$, likely corresponding to Na$_2$CO$_3$. These differences suggest that the type of mineral that forms during freezing depends on the pH of the liquid.

These results demonstrate that Raman spectroscopy can identify carbonate minerals present on the surface of Enceladus. Furthermore, qualitative identification of surface carbonate minerals may enable the estimation of the pH of its subsurface ocean. Therefore, a Raman spectrometer could be an important analytical instrument for in situ observations of surface materials on Enceladus.

\begin{acknowledgments}
We used the large language model (LLM) ChatGPT (OpenAI) \& Gemini (Google) for assistance with language editing, formatting, and general proofreading. This work was supported by the JSPS KAKENHI Grant Number JP23K25921,and JP23H00141.
\end{acknowledgments}

\bibliography{sample701}{}

\begin{thebibliography}{}
\expandafter\ifx\csname natexlab\endcsname\relax\def\natexlab#1{#1}\fi
\providecommand{\url}[1]{\href{#1}{#1}}
\providecommand{\dodoi}[1]{doi:~\href{http://doi.org/#1}{\nolinkurl{#1}}}
\providecommand{\doeprint}[1]{\href{http://ascl.net/#1}{\nolinkurl{http://ascl.net/#1}}}
\providecommand{\doarXiv}[1]{\href{https://arxiv.org/abs/#1}{\nolinkurl{https://arxiv.org/abs/#1}}}

\bibitem[{R.~H. {Brown} {et~al.}(2006){Brown}, {Clark}, {Buratti}, {Cruikshank}, {Barnes}, {Mastrapa}, {Bauer}, {Newman}, {Momary}, {Baines}, {Bellucci}, {Capaccioni}, {Cerroni}, {Combes}, {Coradini}, {Drossart}, {Formisano}, {Jaumann}, {Langevin}, {Sotin}, {et~al.}}]{2006Sci...311.1425B}
{Brown}, R.~H., {Clark}, R.~N., {Buratti}, B.~J., {et~al.} 2006, \bibinfo{title}{{Composition and Physical Properties of Enceladus' Surface},} Science, 311, 1425, \dodoi{10.1126/science.1121031}

\bibitem[{M.~L. {Cable} {et~al.}(2021){Cable}, {Porco}, {Glein}, {German}, {MacKenzie}, {Neveu}, {Hoehler}, {Hofmann}, {Hendrix}, {Eigenbrode}, {Postberg}, {Spilker}, {McEwen}, {Khawaja}, {Waite}, {Wurz}, {Helbert}, {Anbar}, {de Vera}, \& {N{\'u}{\~n}ez}}]{2021PSJ.....2..132C}
{Cable}, M.~L., {Porco}, C., {Glein}, C.~R., {et~al.} 2021, \bibinfo{title}{{The Science Case for a Return to Enceladus},} The Planetary Science Journal, 2, 132, \dodoi{10.3847/PSJ/abfb7a}

\bibitem[{G. {Choblet} {et~al.}(2017){Choblet}, {Tobie}, {Sotin}, {B{\v{e}}hounkov{\'a}}, {{\v{C}}adek}, {Postberg}, \& {Sou{\v{c}}ek}}]{2017NatAs...1..841C}
{Choblet}, G., {Tobie}, G., {Sotin}, C., {et~al.} 2017, \bibinfo{title}{{Powering prolonged hydrothermal activity inside Enceladus},} Nature Astronomy, 1, 841, \dodoi{10.1038/s41550-017-0289-8}

\bibitem[{M. {Choukroun} {et~al.}(2021){Choukroun}, {Backes}, {Cable}, {Fayolle}, {Hodyss}, {Murdza}, {Schulson}, {Badescu}, {Malaska}, {Marteau}, {Molaro}, {Moreland}, {Noell}, {Nordheim}, {Okamoto}, {Riccobono}, \& {Zacny}}]{2021PSJ.....2..100C}
{Choukroun}, M., {Backes}, P., {Cable}, M.~L., {et~al.} 2021, \bibinfo{title}{{Sampling Plume Deposits on Enceladus' Surface to Explore Ocean Materials and Search for Traces of Life or Biosignatures},} The Planetary Science Journal, 2, 100, \dodoi{10.3847/PSJ/abf2c5}

\bibitem[{A.~N. {Deutsch} {et~al.}(2022){Deutsch}, {Panicucci}, {Tenelanda-Osorio}, {Da Poian}, {Cho}, {Venigalla}, {et~al.}}]{2022FrASS...928357D}
{Deutsch}, A.~N., {Panicucci}, P., {Tenelanda-Osorio}, L.~I., {et~al.} 2022, \bibinfo{title}{{The ETNA mission concept: Assessing the habitability of an active ocean world},} Frontiers in Astronomy and Space Sciences, 9, 1028357, \dodoi{10.3389/fspas.2022.1028357}

\bibitem[{L.~M. {Fifer} {et~al.}(2022){Fifer}, {Catling}, \& {Toner}}]{2022PSJ.....3..191F}
{Fifer}, L.~M., {Catling}, D.~C., \& {Toner}, J.~D. 2022, \bibinfo{title}{{Chemical Fractionation Modeling of Plumes Indicates a Gas-rich, Moderately Alkaline Enceladus Ocean},} The Planetary Science Journal, 3, 191, \dodoi{10.3847/PSJ/ac7a9f}

\bibitem[{M.~G. {Fox-Powell} \& C.~R. {Cousins}(2021){Fox-Powell} \& {Cousins}}]{2021JGRE..12660516F}
{Fox-Powell}, M.~G., \& {Cousins}, C.~R. 2021, \bibinfo{title}{{Partitioning of Crystalline and Amorphous Phases During Freezing of Simulated Enceladus Ocean Fluids},} Journal of Geophysical Research: Planets, 126, e06516, \dodoi{10.1029/2020JE006628}

\bibitem[{C.~R. {Glein} {et~al.}(2015){Glein}, {Baross}, \& {Waite}}]{2015GeCoA.162..202G}
{Glein}, C.~R., {Baross}, J.~A., \& {Waite}, J.~H., J. 2015, \bibinfo{title}{{The pH of Enceladus' ocean},} Geochim. Cosmochim. Acta, 162, 202, \dodoi{10.1016/j.gca.2015.04.017}

\bibitem[{C.~R. {Glein} \& N. {Truong}(2025){Glein} \& {Truong}}]{2025Icar..42616717G}
{Glein}, C.~R., \& {Truong}, N. 2025, \bibinfo{title}{{Phosphates reveal high pH ocean water on Enceladus},} Icarus, 116717, \dodoi{10.1016/j.icarus.2025.116717}

\bibitem[{C.~R. {Glein} \& J.~H. {Waite}(2020){Glein} \& {Waite}}]{2020GeRL...4786484G}
{Glein}, C.~R., \& {Waite}, J.~H. 2020, \bibinfo{title}{{The Carbonate Geochemistry of Enceladus' Ocean},} Geophysical Research Letters, 47, e86484, \dodoi{10.1029/2019GL085885}

\bibitem[{B. {Kasalica} {et~al.}(2013){Kasalica}, {Stojadinovi{\'c}}, {Bel{\v{c}}a}, {Sarvan}, {Zekovi{\'c}}, \& {Radi{\'c}-Peri{\'c}}}]{2013JAAS...28...92K}
{Kasalica}, B., {Stojadinovi{\'c}}, S., {Bel{\v{c}}a}, I., {et~al.} 2013, \bibinfo{title}{{The anomalous sodium doublet D 2/D 1 spectral line intensity ratio -- a manifestation of CCD's presaturation effect},} Journal of Analytical Atomic Spectrometry, 28, 92, \dodoi{10.1039/c2ja30239j}

\bibitem[{S. {Kempf} {et~al.}(2008){Kempf}, {Beckmann}, {Moragas-Klostermeyer}, {Postberg}, {Srama}, {Economou}, {Schmidt}, {Spahn}, \& {Gr{\"u}n}}]{2008Icar..193..420K}
{Kempf}, S., {Beckmann}, U., {Moragas-Klostermeyer}, G., {et~al.} 2008, \bibinfo{title}{{The E ring in the vicinity of Enceladus: I. Spatial distribution and properties of the ring particles},} Icarus, 193, 420, \dodoi{10.1016/j.icarus.2007.06.027}

\bibitem[{K. {Konstantinidis} {et~al.}(2015){Konstantinidis}, {Martinez}, {Dachwald}, {Ohndorf}, {Dykta}, {Bowitz}, {et~al.}}]{2015AcAau.106...63K}
{Konstantinidis}, K., {Martinez}, C.~L.~F., {Dachwald}, B., {et~al.} 2015, \bibinfo{title}{{A lander mission to probe subglacial water on Saturn's moon Enceladus for life},} Acta Astronautica, 106, 63, \dodoi{10.1016/j.actaastro.2014.09.012}

\bibitem[{C. {Liu} {et~al.}(2024){Liu}, {Xu}, {Zhang}, {Robinson}, {Lau}, {Huang}, {Glein}, \& {Hao}}]{2024ApJ...971...51L}
{Liu}, C., {Xu}, W., {Zhang}, Z., {et~al.} 2024, \bibinfo{title}{{The Potential for Organic Synthesis in the Ocean of Enceladus},} The Astrophysical Journal, 971, 51, \dodoi{10.3847/1538-4357/ad534f}

\bibitem[{S.~M. {MacKenzie} {et~al.}(2021){MacKenzie}, {Neveu}, {Davila}, {Lunine}, {Craft}, {Cable}, {Phillips-Lander}, {Hofgartner}, {Eigenbrode}, {Waite}, {Glein}, {Gold}, {Greenauer}, {Kirby}, {Bradburne}, {Kounaves}, {Malaska}, {Postberg}, {Patterson}, {et~al.}}]{2021PSJ.....2...77M}
{MacKenzie}, S.~M., {Neveu}, M., {Davila}, A.~F., {et~al.} 2021, \bibinfo{title}{{The Enceladus Orbilander Mission Concept: Balancing Return and Resources in the Search for Life},} The Planetary Science Journal, 2, 77, \dodoi{10.3847/PSJ/abe4da}

\bibitem[{S. {Maurice} {et~al.}(2021){Maurice}, {Wiens}, {Bernardi}, {Cais}, {Robinson}, {Nelson}, {et~al.}}]{2021SSRv..217...47M}
{Maurice}, S., {Wiens}, R.~C., {Bernardi}, P., {et~al.} 2021, \bibinfo{title}{{The SuperCam instrument suite on the Mars 2020 rover: Science objectives and mast-unit description.},} Space Science Reviews, 217, 47, \dodoi{10.1007/s11214-021-00807-w}

\bibitem[{K.~L. {Mitchell} {et~al.}(2024){Mitchell}, {Rabinovitch}, {Scamardella}, \& {Cable}}]{2024JGRE..12908070M}
{Mitchell}, K.~L., {Rabinovitch}, J., {Scamardella}, J.~C., \& {Cable}, M.~L. 2024, \bibinfo{title}{{A Proposed Model for Cryovolcanic Activity on Enceladus Driven by Volatile Exsolution},} Journal of Geophysical Research: Planets, 129, e08070, \dodoi{10.1029/2023JE007977}

\bibitem[{D.~M. {Murphy} \& T. {Koop}(2005){Murphy} \& {Koop}}]{2005QJRMS.131.1539M}
{Murphy}, D.~M., \& {Koop}, T. 2005, \bibinfo{title}{{Review of the vapour pressures of ice and supercooled water for atmospheric applications},} Quarterly Journal of the Royal Meteorological Society, 131, 1539, \dodoi{10.1256/qj.04.94}

\bibitem[{E. {Nathan} {et~al.}(2022){Nathan}, {Balachandran}, {Cappuccio}, {Di}, {Doerksen}, {Gloder}, {et~al.}}]{2022FrASS...995941N}
{Nathan}, E., {Balachandran}, K., {Cappuccio}, P., {et~al.} 2022, \bibinfo{title}{{A multi-lander New Frontiers mission concept study for Enceladus: SILENUS},} Frontiers in Astronomy and Space Sciences, 9, 995941, \dodoi{10.3389/fspas.2022.995941}

\bibitem[{M. {Neveu} {et~al.}(2020){Neveu}, {Anbar}, {Davila}, {Glavin}, {MacKenzie}, {Phillips-Lander}, {Sherwood}, {Takano}, {Williams}, \& {Yano}}]{2020FrASS...7...26N}
{Neveu}, M., {Anbar}, A.~D., {Davila}, A.~F., {et~al.} 2020, \bibinfo{title}{{Returning Samples From Enceladus for Life Detection},} Frontiers in Astronomy and Space Sciences, 7, 26, \dodoi{10.3389/fspas.2020.00026}

\bibitem[{I. {O'Bannon} {et~al.}(2014){O'Bannon}, {Beavers}, \& {Williams}}]{2014AmMin..99.1973O}
{O'Bannon}, E., I., {Beavers}, C.~M., \& {Williams}, Q. 2014, \bibinfo{title}{{Trona at extreme conditions: A pollutant-sequestering material at high pressures and low temperatures},} American Mineralogist, 99, 1973, \dodoi{10.2138/am-2014-4919}

\bibitem[{M.~A. {Pasek}(2020){Pasek}}]{2020FrASS...7...14P}
{Pasek}, M.~A. 2020, \bibinfo{title}{{Plume sample modification at icy moons: implications for biosignatures},} Frontiers in Astronomy and Space Sciences, 7, 14, \dodoi{10.3389/fspas.2020.00014}

\bibitem[{J.~S. {Peter} {et~al.}(2024){Peter}, {Nordheim}, \& {Hand}}]{2024NatAs...8..164P}
{Peter}, J.~S., {Nordheim}, T.~A., \& {Hand}, K.~P. 2024, \bibinfo{title}{{Detection of HCN and diverse redox chemistry in the plume of Enceladus},} Nature Astronomy, 8, 164, \dodoi{10.1038/s41550-023-02160-0}

\bibitem[{C.~M. {Phillips-Lander} {et~al.}(2019){Phillips-Lander}, {Moore}, {Raut}, {Molyneaux}, {Miller}, {Nowicki}, {Blase}, {Davis}, {Veach}, {Dirks}, {Persson}, {Tyler}, {Klar}, {Karnes}, {Freeman}, {Howett}, {Soto}, {Mandt}, {Roth}, \& {Retherford}}]{2019LPI....50.2992P}
{Phillips-Lander}, C.~M., {Moore}, T.~Z., {Raut}, U., {et~al.} 2019, \bibinfo{title}{{Europa Integrating Cavity Enhanced Raman Spectrometer for Exploration of Icy Worlds (ERSO) Concept},} in Lunar and Planetary Science Conference, 2992

\bibitem[{C.~C. {Porco} {et~al.}(2006){Porco}, {Helfenstein}, {Thomas}, {Ingersoll}, {Wisdom}, {West}, {Neukum}, {Denk}, {Wagner}, {Roatsch}, {Kieffer}, {Turtle}, {McEwen}, {Johnson}, {Rathbun}, {Veverka}, {Wilson}, {Perry}, {Spitale}, {et~al.}}]{2006Sci...311.1393P}
{Porco}, C.~C., {Helfenstein}, P., {Thomas}, P.~C., {et~al.} 2006, \bibinfo{title}{{Cassini Observes the Active South Pole of Enceladus},} Science, 311, 1393, \dodoi{10.1126/science.1123013}

\bibitem[{F. {Postberg} {et~al.}(2018){Postberg}, {Clark}, {Hansen}, {Coates}, {Dalle Ore}, {Scipioni}, {Hedman}, \& {Waite}}]{2018eims.book..129P}
{Postberg}, F., {Clark}, R.~N., {Hansen}, C.~J., {et~al.} 2018, \bibinfo{title}{{Plume and surface composition of Enceladus},} in Enceladus and the Icy Moons of Saturn, ed. P.~M. {Schenk}, R.~N. {Clark}, C.~J.~A. {Howett}, A.~J. {Verbiscer}, \& J.~H. {Waite} (University of Arizona Press), 129--162, \dodoi{10.2458/azu_uapress_9780816537075-ch007}

\bibitem[{F. {Postberg} {et~al.}(2009){Postberg}, {Kempf}, {Schmidt}, {Brilliantov}, {Beinsen}, {Abel}, {Buck}, \& {Srama}}]{2009Natur.459.1098P}
{Postberg}, F., {Kempf}, S., {Schmidt}, J., {et~al.} 2009, \bibinfo{title}{{Sodium salts in E-ring ice grains from an ocean below the surface of Enceladus},} Nature, 459, 1098, \dodoi{10.1038/nature08046}

\bibitem[{F. {Postberg} {et~al.}(2011){Postberg}, {Schmidt}, {Hillier}, {Kempf}, \& {Srama}}]{2011Natur.474..620P}
{Postberg}, F., {Schmidt}, J., {Hillier}, J., {Kempf}, S., \& {Srama}, R. 2011, \bibinfo{title}{{A salt-water reservoir as the source of a compositionally stratified plume on Enceladus},} Nature, 474, 620, \dodoi{10.1038/nature10175}

\bibitem[{J. {Schmidt} {et~al.}(2008){Schmidt}, {Brilliantov}, {Spahn}, \& {Kempf}}]{2008Natur.451..685S}
{Schmidt}, J., {Brilliantov}, N., {Spahn}, F., \& {Kempf}, S. 2008, \bibinfo{title}{{Slow dust in Enceladus' plume from condensation and wall collisions in tiger stripe fractures},} Nature, 451, 685, \dodoi{10.1038/nature06491}

\bibitem[{L. {Schwander} {et~al.}(2023){Schwander}, {Brabender}, {Mrnjavac}, {Wimmer}, {Preiner}, \& {Martin}}]{2023FrMic..1457597S}
{Schwander}, L., {Brabender}, M., {Mrnjavac}, N., {et~al.} 2023, \bibinfo{title}{{Serpentinization as the source of energy, electrons, organics, catalysts, nutrients and pH gradients for the origin of LUCA and life},} Frontiers in Microbiology, 14, 1257597, \dodoi{10.3389/fmicb.2023.1257597}

\bibitem[{S. {Sharma} {et~al.}(2023){Sharma}, {Roppel}, {Murphy}, {Beegle}, {Bhartia}, {Steele}, {et~al.}}]{2023Natur.619..724S}
{Sharma}, S., {Roppel}, R.~D., {Murphy}, A.~E., {et~al.} 2023, \bibinfo{title}{{Diverse organic-mineral associations in Jezero crater, Mars},} Nature, 619, 724, \dodoi{10.1038/s41586-023-06143-z}

\bibitem[{S.~K. {Sharma} {et~al.}(2020){Sharma}, {Porter}, {Misra}, {Acosta-Maeda}, {Angel}, \& {McKay}}]{2020JRS....51.1782S}
{Sharma}, S.~K., {Porter}, J.~N., {Misra}, A.~K., {et~al.} 2020, \bibinfo{title}{{Standoff Raman spectroscopy for future Europa Lander missions},} Journal of Raman Spectroscopy, 51, 1782, \dodoi{10.1002/jrs.5814}

\bibitem[{J.~R. {Spencer} {et~al.}(2006){Spencer}, {Pearl}, {Segura}, {Flasar}, {Mamoutkine}, {Romani}, {Buratti}, {Hendrix}, {Spilker}, \& {Lopes}}]{2006Sci...311.1401S}
{Spencer}, J.~R., {Pearl}, J.~C., {Segura}, M., {et~al.} 2006, \bibinfo{title}{{Cassini Encounters Enceladus: Background and the Discovery of a South Polar Hot Spot},} Science, 311, 1401, \dodoi{10.1126/science.1121661}

\bibitem[{J.~R. {Spencer} {et~al.}(2009){Spencer}, {Barr}, {Esposito}, {Helfenstein}, {Ingersoll}, {Jaumann}, {McKay}, {Nimmo}, \& {Waite}}]{2009sfch.book..683S}
{Spencer}, J.~R., {Barr}, A.~C., {Esposito}, L.~W., {et~al.} 2009, \bibinfo{title}{{Enceladus: An Active Cryovolcanic Satellite},} in Saturn from Cassini-Huygens (Springer), 683--724, \dodoi{10.1007/978-1-4020-9217-6_21}

\bibitem[{P.~C. {Thomas} {et~al.}(2016){Thomas}, {Tajeddine}, {Tiscareno}, {Burns}, {Joseph}, {Loredo}, {Helfenstein}, \& {Porco}}]{2016Icar..264...37T}
{Thomas}, P.~C., {Tajeddine}, R., {Tiscareno}, M.~S., {et~al.} 2016, \bibinfo{title}{{Enceladus's measured physical libration requires a global subsurface ocean},} Icarus, 264, 37, \dodoi{10.1016/j.icarus.2015.08.037}

\bibitem[{T.~H. {Vu} {et~al.}(2023){Vu}, {Hodyss}, {Johnson}, \& {Cable}}]{2023PSJ.....4..156V}
{Vu}, T.~H., {Hodyss}, R., {Johnson}, P.~V., \& {Cable}, M.~L. 2023, \bibinfo{title}{{Spatial Distribution of Glycine and Aspartic Acid in Rapidly Frozen Brines Relevant to Enceladus},} The Planetary Science Journal, 4, 156, \dodoi{10.3847/PSJ/aced90}

\bibitem[{J.~H. {Waite} {et~al.}(2017){Waite}, {Glein}, {Perryman}, {Teolis}, {Magee}, {Miller}, {Grimes}, {Perry}, {Miller}, {Bouquet}, {Lunine}, {Brockwell}, \& {Bolton}}]{2017Sci...356..155W}
{Waite}, J.~H., {Glein}, C.~R., {Perryman}, R.~S., {et~al.} 2017, \bibinfo{title}{{Cassini finds molecular hydrogen in the Enceladus plume: Evidence for hydrothermal processes},} Science, 356, 155, \dodoi{10.1126/science.aai8703}

\bibitem[{R.~C. {Wiens} {et~al.}(2021){Wiens}, {Maurice}, {Robinson}, {Nelson}, {Cais}, {Bernardi}, {Newell}, {Clegg}, {Sharma}, {Storms}, {Deming}, {Beckman}, {Ollila}, {Gasnault}, {Anderson}, {Andr{\'e}}, {Michael Angel}, {Arana}, {Auden}, {Willis}, {et~al.}}]{2021SSRv..217....4W}
{Wiens}, R.~C., {Maurice}, S., {Robinson}, S.~H., {et~al.} 2021, \bibinfo{title}{{The SuperCam Instrument Suite on the NASA Mars 2020 Rover: Body Unit and Combined System Tests},} Space Science Reviews, 217, 4, \dodoi{10.1007/s11214-020-00777-5}

\end{thebibliography}
\bibliographystyle{aasjournalv7}



\end{document}